\documentclass[a4paper,floatfix]{jpconf}

\usepackage{graphicx}
\usepackage{gensymb}
\usepackage{amsmath,amssymb,color,hyperref}

\providecommand{\pp}{p-p}
\providecommand{\gaga}{\gamma\,\gamma}
\providecommand{\epem}{\rm e^{+}e^{-}}
\providecommand{\FCCee}{FCC-ee}
\providecommand{\pythia}{{\sc pythia}}
\providecommand{\MG}{{\sc MadGraph}}

\newcommand{\sqrts}{\sqrt{s}}

\newcommand{\bbar}     {\ensuremath{b\bar{b}}}

\newcommand{\qqbar}    {\ensuremath{q\bar{q}}}
\newcommand{\ccbar}    {\ensuremath{c\bar{c}}}

\newcommand{\Lumi}{\mathcal{L}}

\begin{document}


\title{Prospects for $\gaga\to H$ and $\gaga\to W^{+}W^{-}$ measurements at the \FCCee}

\author{Patricia Rebello Teles}
\address{Centro Brasileiro de Pesquisas Fisicas - CBPF\\
22290-180 Rio de Janeiro, Brazil}
\ead{patricia.rebello.teles@cern.ch}

\author{David d'Enterria}
\address{CERN, PH Department\\
1211 Geneva, Switzerland}
\ead{dde@cern.ch}

\begin{abstract}
We study the possibilities for the measurement of two-photon production of the Higgs boson (in the $\bbar$
decay channel), and of $W^{+}W^{-}$ pairs (decaying into four jets) in $\epem$ collisions at the the Future
Circular Collider (\FCCee). The processes are simulated with the \pythia\ and \MG~5 Monte Carlo codes, 
using the  effective photon approximation for the $\epem$ photon fluxes, 
at center-of-mass energies $\sqrt{s} = $~160~GeV and 240~GeV. The analyses include electron-positron tagging, realistic
acceptance and reconstruction efficiencies for the final-state jets, 
and selection criteria to remove the backgrounds. 
Observation of both channels is achievable with the expected few ab$^{-1}$ integrated luminosities at \FCCee.
\end{abstract}

\section{Introduction}
\medskip

After the LHC discovery of the Higgs boson with properties consistent with the  standard model
(SM) expectations~\cite{higgs}, the priority in high-energy particle physics is more-than-ever focused on 
looking for evidences of ``new physics'' that can help explain the many fundamental problems still open in the
field (nature of dark matter, matter-antimatter asymmetry, unnatural gap between the electroweak and Planck
scales,...). At CERN, the design study of the Future Circular Collider (FCC) has been launched to pursue the
searches of physics beyond the SM in a new 80--100~km tunnel once the LHC research project is
completed~\cite{fcchh}. Running in its first phase as a very-high-luminosity electron-positron collider, the \FCCee\
(formerly known as TLEP~\cite{tlep}) will provide unique possibilities for indirect searches of new phenomena
through high-precision tests of the SM by collecting tens of ab$^{-1}$ integrated luminosities at the $Z$ pole
($\sqrt{s}=91$~GeV), the $W$-pair production threshold ($\sqrt{s}=160$~GeV), the $HZ$ Higgs-strahlung maximum
around $\sqrt{s}=240$~GeV, and the $t\bar{t}$ threshold ($\sqrt{s}=350$~GeV).\\

Beyond the rich physics programme in electron-positron collisions, the \FCCee\ will provide an ideal
environment to study $\gaga$ collisions at unprecedented energies and luminosities.
Though photon-fusion processes have been usually considered at dedicated high-energy photon-colliders through
Compton-backscattering of laser light~\cite{Telnov:1995hc}, they can also be studied using the photons
radiated from the $\epem$ beams. Indeed, any relativistic charged particle generates a flux of quasi-real
(Weizs\"acker-Williams) photons, theoretically described using the effective photon approximation
(EPA)~\cite{epa}, which can  be used for photoproduction studies. The physics of two-photon collisions
has been an active topic of research at $\epem$, proton, and ion colliders (see
e.g.~\cite{d'Enterria:2008zz,Baltz:2007kq} and refs. therein). 
The $\gaga$ kinematics can be constrained measuring the scattered leptons with near-beam detectors at
low angles, while the produced system is reconstructed from its decaying products in the central detector.
In this work, we are interested, first, in the $\gaga\to H$ process, considered 
before in the context of photon~\cite{Brodsky:1994nf}, proton~\cite{Khoze:2001xm} and
heavy-ion~\cite{david} colliders. The huge luminosities available at \FCCee\ provide the means to observe
such a process which, to our knowledge, has not been studied before in $\epem$ two-photon EPA collisions.
The second measurement of interest is $\gaga\to W^{+}W^{-}$, which is a very suitable process to test the
electroweak sector of the SM as it probes trilinear $W^{+}W^{-}\gamma$ and quartic $\gaga W^{+}W^{-}$ boson
couplings. 
The few counts of two-photon production of $W^{+}W^{-}$ observed for the first time at the
LHC~\cite{Chatrchyan:2013akv,CMSaaww} have improved the limits on anomalous quartic gauge couplings (aQGC)
using dimension- 6 and 8 effective operators parameters.
The cleaner environment and huge luminosities of $\epem$ collisions at the \FCCee\ will lead to hundreds
of WW pairs observed and, thus, much more stringent aQGC limits.

\begin{figure}[htbp!]
  \centering
  \includegraphics[width=0.8\columnwidth]{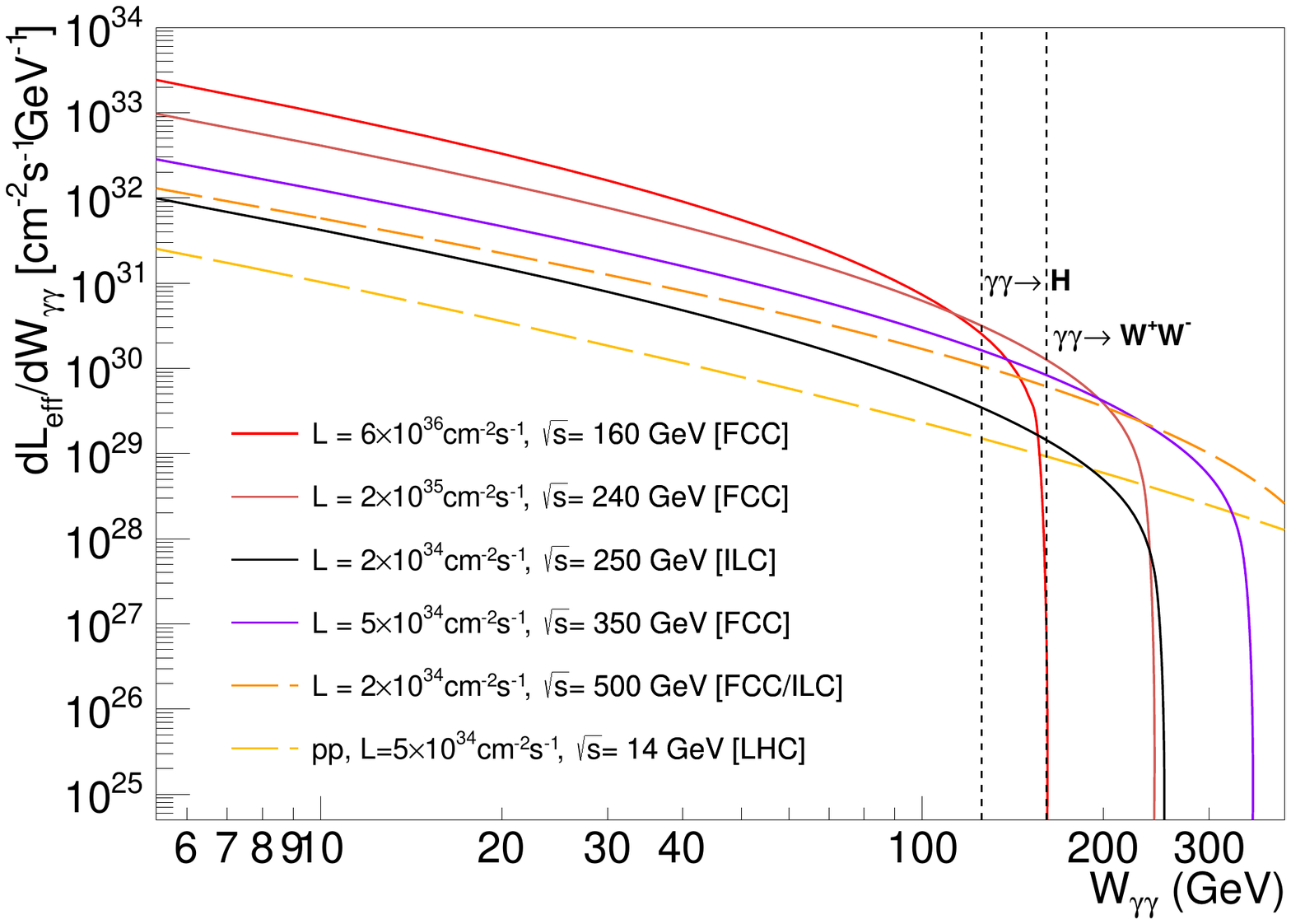}
  \caption{\label{lumifcc} Effective two-photon luminosities as a function of $\gaga$ c.m. energy,
  from the EPA fluxes convolution at \FCCee\ and ILC, 
  and \pp\ at the LHC, for their nominal beam luminosities. The dashed vertical lines show the 
  $\gaga \to H$ and $\gaga\to W^+W^-$ production thresholds.}
\end{figure}

Figure~\ref{lumifcc} shows the $\gaga$ effective luminosities $\Lumi_{\rm eff}$ obtained from EPA fluxes for
lepton and hadron colliders at various center-of-mass (c.m.) energies. The vertical dashed lines indicate the
threshold of production for the two processes considered here. Photon-fusion Higgs production can be in principle
observed at \FCCee\ running at $\sqrts$~=~160~GeV, although the best 
conditions for measuring both processes are provided by $\epem$ at $\sqrts=240$~GeV. In this case, the 
$\Lumi_{\rm eff}$ is 20 times greater than at the LHC, with the advantage of the absence of pile-up collisions.  


\section{Theoretical setup}
\medskip

\begin{figure}[htbp!]
  \centering
\includegraphics[width=32pc]{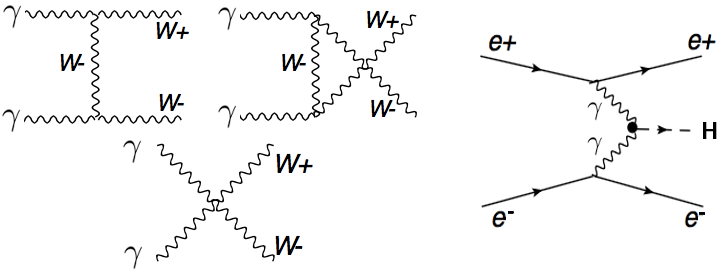}
\caption{\label{diagrams} Diagrams for $W^{+}W^{-}$ (left) and Higgs boson (right) $\gamma$-fusion production in $\epem$ collisions.} 
\end{figure}

All charges accelerated at high energies generate electromagnetic fields which, in the EPA
approach, can be considered as quasi-real (low virtuality) photon beams. After interacting with each other,
the two photons can produce various exclusive final states 
$\epem \to e^{+} \, \gaga \,e^{-} \to e^{+}\;$$X$$\;e^{-}$, where $X$ is either $W^{+}W^{-}$ or the Higgs
boson in this work (diagrams in Fig.~\ref{diagrams}), and the radiated leptons are scattered at very small
angles with respect to the beam direction. The production cross sections are computed in a factorized way
through the convolution of the EPA photon fluxes and the elementary $\gaga\to X$ process described by exact
matrix elements and kinematics. Although various Monte Carlo (MC) event generators --such as \pythia~6~\cite{pythia6},
\MG~5~\cite{madgraph}, {\sc whizard}~\cite{whizard} or {\sc SuperChic}~2~\cite{chic}-- have in principle the
possibility to run $\gaga$ collisions in the EPA mode, not all production processes are available and/or the
simulation of the outgoing $\epem$ kinematics is not always exact (and such information is crucial for
realistic studies of the $\epem$ tagging probability). In this study, we employ two different MC codes, \MG~5
(version 2.1.1) and \pythia~6 (version 6.4.28), to study respectively Higgs and  $W^{+}W^{-}$ two-photon production.\\

We run \MG~5 with the effective field theory (HEFT) approach, where the scalar Higgs couples directly to
photons, combined with the EPA to compute the cross sections for the process 
$\gaga \to H \to b \bar{b}$, and the relevant continuum
backgrounds $\gaga \to \bbar,\ccbar,\qqbar$ (with $q=u,d,s$ being light quarks).
The final-states are studied at the parton level, i.e. neither quark showering nor hadronization and 
decays are considered for this preliminary study (a more complete analysis is in progress).
The simulated signal and background samples for $\gaga \to W^{+}W^{-}\to 4$~jets were generated using
\pythia~6~\cite{pythia6} in the EPA approach (process ID=69, which properly accounts for the scattered
$e^{+}e^{-}$ kinematics) and full parton-shower and hadronization of the jets from the W decays.
With both gauge bosons $W^{\pm}$ producing four jets in the final state, the statistics
is maximized due to the higher branching ratio compared to the leptonic decay channels used in the CMS
analyses in \pp\ at 7 and 8~TeV~\cite{Chatrchyan:2013akv,CMSaaww}.



\section{Results}
\medskip


\subsection{$\gaga \to H \to\bbar$ process}
\medskip

\begin{figure}[htbp!]
  \centering
  \includegraphics[width=35pc]{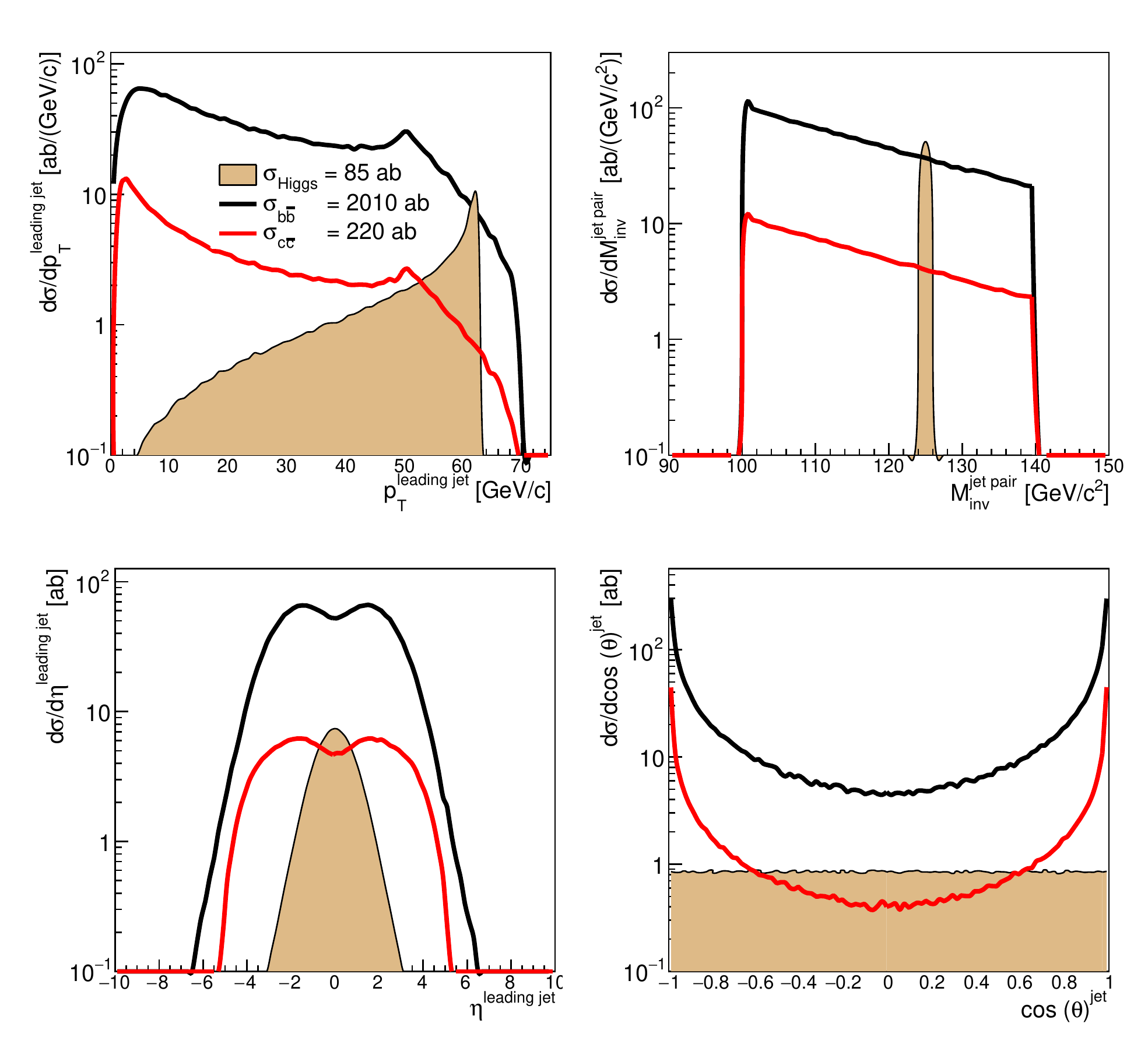}
  \caption{\label{higgsnocuts}Kinematical distributions of the $b$-jets from signal (filled histograms) and backgrounds in the
    analysis $\gaga \to H\to\bbar$ at $\sqrt{s} = 240$~GeV: transverse momentum $p_{T}^{j}$ (top left) and
    pseudorapidity $\eta^{j}$ (bottom left) of each jet; invariant mass $M_{jj}$ of the jet pair (top right), and
    cosine of the angle $\theta$ between the jet, boosted to the rest frame of the pair, and the
    direction of the pair (helicity frame, bottom right). All distributions are normalized to their visible cross sections
    assuming the $b$-jet reconstruction and (mis)tagging efficiencies quoted in the text.}
\end{figure}

The visible cross sections for $\gaga \to H \to\bbar$ obtained with \MG~5 are $\sigma_{\gaga \to H\to\bbar}$~=~20~ab
and 85~ab in $\epem$ collisions at $\sqrt{s} = $160~GeV and 240~GeV respectively, assuming a 70\%
reconstruction/tagging efficiency for each single $b$-jet (which leads to a combined 50\% efficiency for the
$\bbar$ signal). The main backgrounds to the signal are the continuums $\gaga\to\bbar$, and $\gaga\to\ccbar,\qqbar$
where the charm and light quarks are misidentified as $b$-quarks. Assuming
$b$-jet mistagging probabilities of 5\% for a $c$-quark, and 1.5\% for a light-quark, effective reductions of the
$\ccbar$ and $\qqbar$ continuum backgrounds by factors of about 400 and $4 \times 10^{3}$, respectively, are
achieved. Taking all these factors into account, the backgrounds are still about 30 times larger than the
signal in the region of jet-pair invariant masses $100\;\mbox{GeV} \leq M_{inv}^{jj} \leq 140\;\mbox{GeV}$
around the Higgs peak, but can be safely reduced with a few kinematics cuts that are described below in
detail and summarized in Table~\ref{table1}. Figure~\ref{higgsnocuts} compares the $b$-jets kinematical
distributions for the signal and backgrounds for $\epem$ collisions at $\sqrt{s} = 240$~GeV. The
kinematical distributions for $\sqrt{s} = 160$~GeV are similar and are not shown here for
obvious reasons.

\begin{table}[htbp!]
\caption{\label{table1} Summary of the visible cross sections for signal and backgrounds for the $\gaga \to H \to
  \bbar$ analysis, obtained from the \MG~5 simulations for $\epem$  collisions at $\sqrt{s}$~=~160,~240~GeV,
  following the selection criteria described in the text.} 
\centering
\small{\begin{tabular}{l|c|c|c|c}
\br
Process & $\sqrt{s}$ & $[M^{jj}_{\rm inv}=$~100--140~GeV/c$^{2}]$ & ...                                     & ...          \\
        & (GeV)   &                   ($b$-jet (mis)tag efficiency)                      & ...                              & ...          \\
        &         &  & $[p_{T}^{j}=$~53.5--62.5~GeV/c]     & ...           \\
        &         &  & $|\cos \theta^{j} | < 0.45$              &  ...         \\
        &         &  &  $[M^{\rm jj}_{\rm inv}=$~115--135~GeV/c$^{2}]$ & ...           \\
        &         &  &                                          & $e^{\pm}$-tag   \\\br
$\gaga \to H \to b\bar{b}$  & 160      & 40 ab (20 ab)       & 9.2 ab  & 4.6 ab     \\
$\gaga \to b\bar{b}$        &          & 1\,120 ab (556 ab)  & 16.2 ab  & 8.1 ab     \\
$\gaga \to c\bar{c}$        &          & 21\,610 ab (54 ab)  & 0.13 ab  & 0.06 ab    \\
$\gaga \to q\bar{q}$        &          & 24\,320 ab (5.5 ab) & 0.002 ab & 0.58e-3 ab \\\br
$\gaga \to H \to b\bar{b}$ & 240       & 170 ab (85 ab)     & 39.3 ab & 19.6  ab     \\
$\gaga \to b\bar{b}$        &          & 4\,020 ab (2010 ab) & 67.8 ab &  34.0 ab           \\
$\gaga \to c\bar{c}$        &          & 77\,830 ab (195 ab) & 6.0 ab  &  3.0 ab             \\
$\gaga \to q\bar{q}$        &          & 87\,760 ab (20 ab)  & 0.005 ab &  0.002 ab          \\
\br
  \end{tabular}
}
\end{table}

\begin{figure}[htbp!]
  \centering
  \includegraphics[width=35pc]{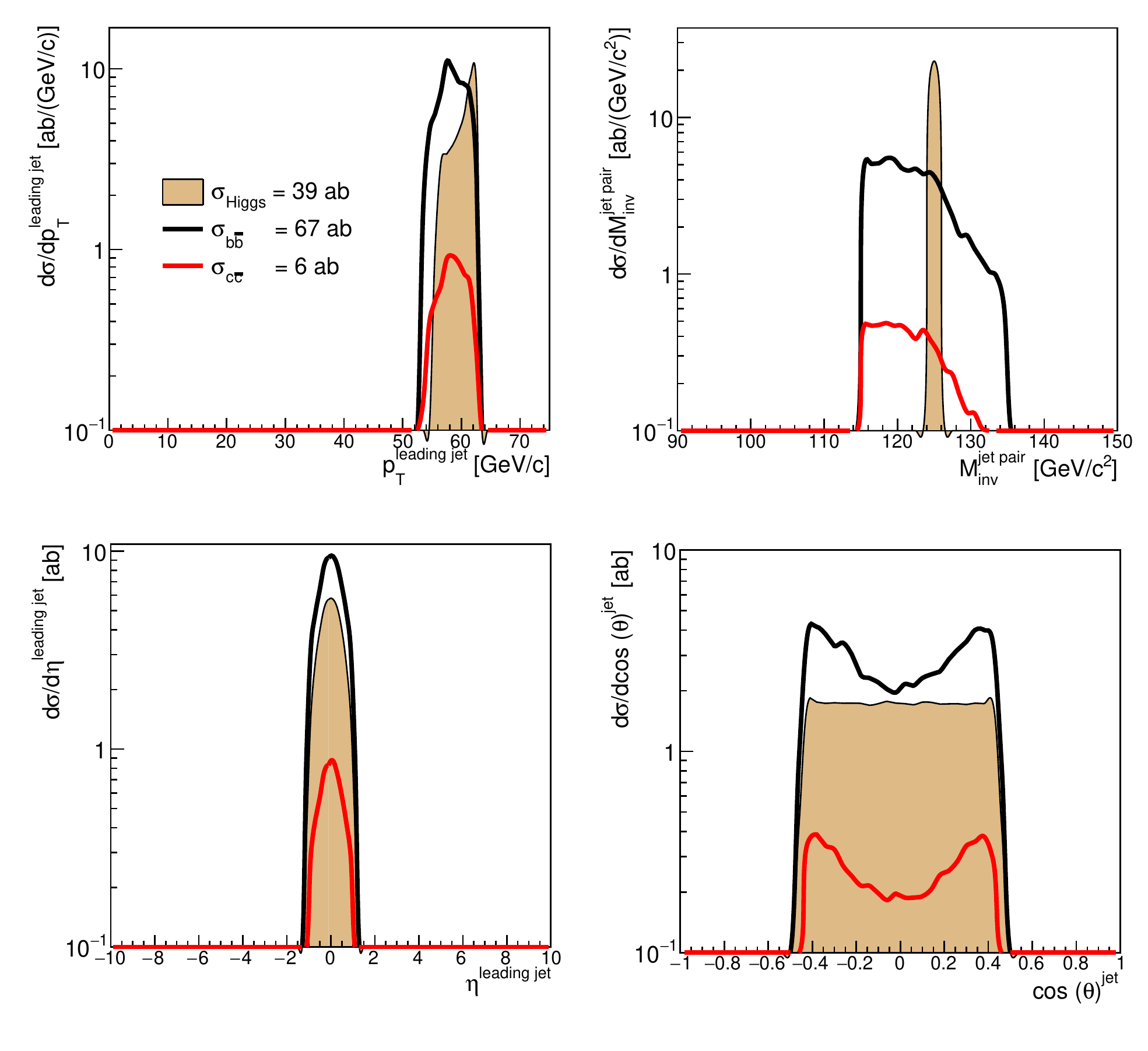}
  \caption{\label{higgscuts}Kinematical distributions of the $b$-jets from signal (filled histograms) and backgrounds in the
    analysis $\gaga \to H\to\bbar$ at $\sqrt{s} = 240$~GeV: transverse momentum $p_{T}^{j}$ (top left) and
    pseudorapidity $\eta^{j}$ (bottom left) of each jet, invariant mass $M_{jj}$ of the jet pair (top right),
    and cosine of the angle $\theta$ between the jet, boosted to the rest frame of the pair, and the
    direction of the pair (helicity frame, bottom right); after application of all selection criteria mentioned in
    the text.} 
  \end{figure}

In order to remove the continuum backgrounds, several cuts can be applied according to the study~\cite{david} and the
distributions shown in Fig.~\ref{higgsnocuts}. First, since the transverse momenta of the Higgs decay $b$-jets peak at 
$p_{T}^{j}\approx m_{_{\rm H}}/2$, selecting events with at least one jet within $p_{T}=$~53.5--62.5 GeV/c
suppresses $\sim$95\% of the backgrounds, while removing only $\sim$50\% of the signal. Second, the two
decay $b$-jets are emitted isotropically in the Higgs c.m. system, and their distribution in the helicity frame is
peaked at $| \cos \theta |\approx 1$, i.e. emitted either roughly in the same direction as the $\bbar$ pair
or opposite to it, while the continuum --whose relevant Feynman diagrams have quarks propagating in the t- or
u- channel-- is peaked in the forward and backward directions. Thus, independently requiring $|\cos\theta|<0.45$ 
suppresses $\sim$90\% of the continuum contaminations while still keeping $\sim$55\% of
signal. Combination of both criteria enhances the (S)ignal/(B)ackground by a factor of $\sim$15
(if needed one could further exploit the jet pseudorapidity distributions, which are peaked around
$\eta\approx$~0 for the signal and more spread out for the continuum).
The final Higgs yield extraction is obtained integrating the invariant mass distribution of the jet
pairs within $M^{jj}_{inv}=$~115--135~GeV/c$^{2}$, i.e. $\pm$2$\sigma$ around the assumed $\sim$3.5~GeV
per-jet energy resolution. This has no effect on the final signal but further reduces 69\% of the
backgrounds below the (now narrower) peak region considered. 
It's worth mentioning that we didn't consider any non-$\gaga$ backgrounds from $\epem\to\bbar$
processes. Our assumption is that all such events are removed by requiring a double-tagging condition on
opposite-sign electron/positron at each side of the detector, with kinematics consistent with the
reconstructed central system. %
We assume e$^\pm$ detection over $1^{\degree} < \theta_{e^{\pm}} < 179^{\degree}$, where
$\theta_{e^{\pm}}$ is the angle of the scattered $e^{\pm}$ with respect to the beam, and 50\% of
efficiency for the double-tagging of $\gaga$ events.
The visible cross sections remaining after application of the aforementioned selection cuts are listed in 
Table~\ref{table1}, and their impact in the final kinematical distributions are shown in Fig.~\ref{higgscuts}.
From the final visible cross sections listed in the fifth column of Table~\ref{table1}, we expect evidence
(S/$\sqrt{B}$~$>$~3$\sigma$) for Higgs boson production in the $b\bar{b}$ decay channel at $\sqrt{s} = 160$~GeV (240~GeV)
with $\Lumi = 3.5\;\mbox{ab}^{-1}$ ($0.9\;\mbox{ab}^{-1}$). Full 5$\sigma$-observation of two-photon Higgs production
in this decay channel would occur at $\sqrt{s} = 160$~GeV (240~GeV) for 
$\Lumi = 10\;\mbox{ab}^{-1}$ ($2.5\;\mbox{ab}^{-1}$). Both observations are thus guaranteed with the
current \FCCee\ plans to integrate  $\Lumi = 15.2\;\mbox{ab}^{-1}$ and $\Lumi = 3.5\;\mbox{ab}^{-1}$ per-year
at each c.m. energy in a total of four interaction points.\\

\subsection{$\gaga \to W^+W^- \to 4$-jets process}
\medskip

\begin{table}[htbp!]
\caption{\label{tableaaww}  Summary of the cross sections values for signal and backgrounds for the $\gaga \to
  W^+W^- \to$~4-jets analysis, obtained from \pythia~6 simulations for $\epem$  at  $\sqrt{s}$~=~240~GeV,
  following the selection criteria described in the text.}
\centering
\small{\begin{tabular}{l|c|c|c|c}
\br
Process & $\sqrt{s}$  & Generator-level (no cut) &   $p_{T}^{j} > 10 \mbox{GeV/c}$     & ...           \\
        &                  &                                         & $|\eta^{j}| < 0.5$              &  ...         \\
        &                  &                                         &  $\Delta R^{jj} > 0.4$ & ...           \\
        &                  &                                         &                          & $[M^{jj}=$ 76.5--84.5~GeV/c$^{2}]$ \\
        &                  &                                         &                                          & $e^{\pm}$-tag   \\
\br
$\gamma \gamma \to W^{+}W^{-} \to 4 \mbox{jets}$  & 240  &  7\,234 ab   &    5\,809 ab  & 626 ab        \\
$\gamma \gamma \to 4 \mbox{jets}$                 & 240  & 24\,890 ab   &     999.6 ab  & 56 ab          \\
\br
  \end{tabular}
}
\end{table}

\begin{figure}[htbp!]
  \centering
\includegraphics[width=35pc]{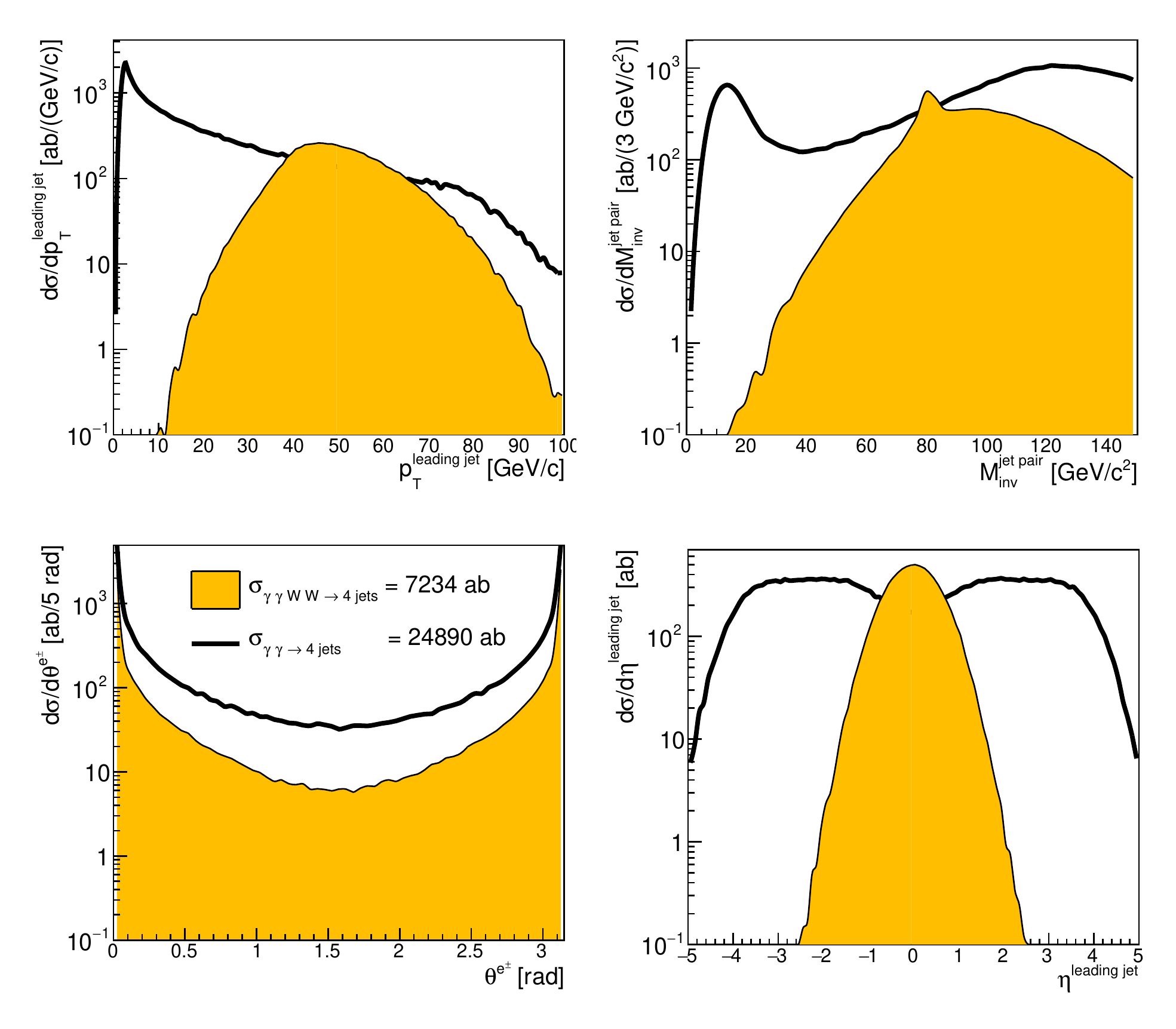}
\caption{\label{aaww240}  Kinematical distributions of signal (filled histograms) and backgrounds in the
    analysis $\gaga \to W^+W^-\to$~4-jets at $\sqrt{s} = 240$~GeV: transverse momentum (top left) and
    pseudorapidity (bottom right) of the leading jet, invariant mass $M_{jj}$ of jet pairs (top right) and
    scattered angle of the outgoing leptons (bottom left), without any selection criteria applied.}
\end{figure}

The generation of the two-photon production of $W^{+}W^{-}$ pairs in the fully hadronic decay channel was
performed with \pythia~6, which predicts a cross section of $\sigma_{\gaga\to WW}$~=~7.23~fb in $\epem$ at
$\sqrts$~=~240~GeV. The dominant background from the process, $e^{+}e^{-}\to 4$~jets with very large cross
sections of $\sigma \approx$~8.8~pb, should be killed with the $e^{\pm}$ double-tagging condition, leaving only
the irreducible $\gaga \to 4$~jets process with cross section $\sigma$~=~25~fb, to deal with.
The $\epem$ $k_T$ (``Durham'') jet reconstruction algorithm~\cite{durham} from the Fastjet
package~\cite{fastjet} has been used to perform the clustering of all the hadronic energy in the event into four exclusive jets.
The transverse momentum $p_{T}$ and pseudorapidity $\eta$ of the leading jet, the invariant mass $M_{jj}$ of
jet pairs and the scattered angle of the outgoing leptons $\theta_{e^{\pm}}$ with respect to the beam are
used as kinematical variables for improving the signal significance against the
background. Figure~\ref{aaww240} compares such kinematical distributions for signal and irreducible background
at $\sqrt{s} = 240$~GeV, before any analysis selection applied.

\begin{figure}[hbp!]
  \centering
\includegraphics[width=35pc]{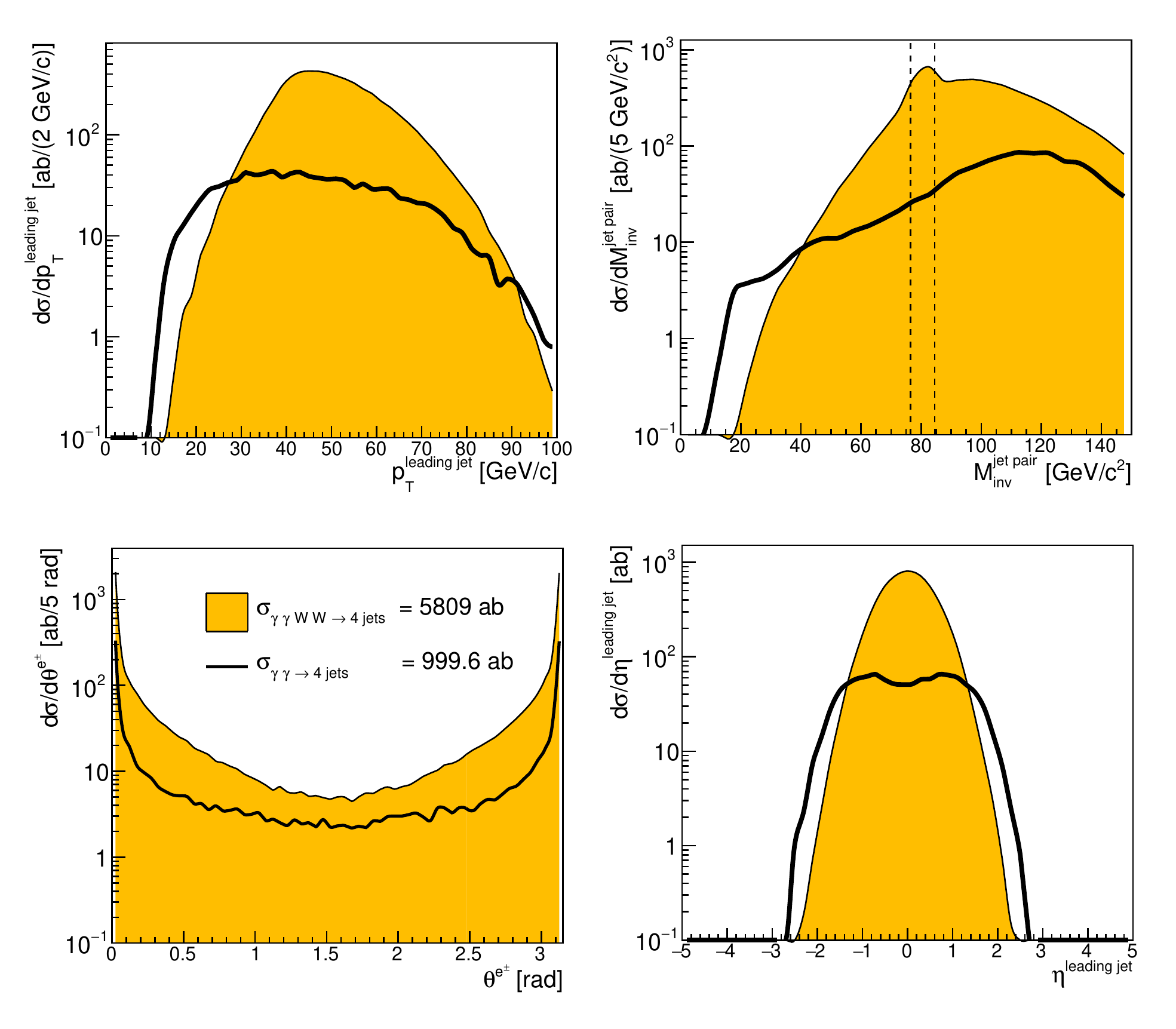}
\caption{\label{aaww240CUTS} Kinematical distributions of signal (filled histograms) and backgrounds in the
    analysis $\gaga \to W^+W^-\to$~4-jets at $\sqrt{s} = 240$~GeV after applying the acceptance cuts
   ($p_{T}^{j} > 10$~GeV, $|\eta^{j}| < 5$ and $\Delta R^{jj} > 0.4$): transverse momentum $p_{T}^{j}$ (top
   left) and pseudorapidity $\eta^{j}$ (bottom right) of the leading jet; invariant mass $M_{jj}$ of the jet
   pair (top right), and  scattered angle of the outgoing leptons (bottom left).} 
\end{figure}

Assuming basic detector acceptance cuts: four exclusive jets reconstructed with $p_{T}^{j} > 10$~GeV,
$|\eta^{j}| < 5$ and $\Delta R^{jj} > 0.4$, 
reduces $\sim$90\% of the continuum background against only $\sim$20\% of the signal. The corresponding
kinematical distributions are plotted in Figure~\ref{aaww240CUTS}. In addition, we assume the tagging of both 
scattered leptons, within the polar angle acceptance range $1^{\degree} \leq \theta_{e^{\pm}} \leq
179^{\degree}$, with a 50\% efficiency. 
The last selection criterion required for improving our signal to background ratio, is an invariant mass of
the jet pairs around the $W^{\pm}$ mass, i.e. within the  76.5~GeV~$\leq M_{jj} \leq$~84.5~GeV mass window.
The combined application of all selection cuts leads to a suppression of a factor of $\sim$10 of the signal
events, and a $\times$450 reduction of the irreducible background. 
For an integrated luminosity of $\Lumi = 1\;\mbox{ab}^{-1}$, we expect more than 600 events for the signal,
i.e. 10 times above the number of irreducible background counts, reaching a signal-to-background ratio of 
$S_{\gamma \gamma \to WW \to 4 jets}/B_{\gaga \to 4 jets} \approx 11$, and a statistical significance of
${\cal S} \approx 25$ obtained with a method based on a profile likelihood ratio~\cite{cowan}.
As a matter of fact, the simple (5$\sigma$) observation of $\gaga \to WW \to 4$~jets  would just require  
$\Lumi = 0.05\;\mbox{ab}^{-1}$ at $\sqrt{s} = 240$~GeV, although strict aQGC tests will profit from the much
larger statistical signal expected with the full integrated luminosity. 



\section{Conclusions}
\medskip

We have presented feasibility studies for the observation of two-photon production of the Higgs boson 
(in the $b\bar{b}$ decay channel) as well as $W^{+}W^{-}$ pairs (in their fully-hadronic decay mode) in
$\epem$ collisions at the \FCCee, using the equivalent photon flux of the colliding beams. Both final-states
are otherwise inaccessible at the LHC due to the huge backgrounds in their  
full-jet decay channels. Results are presented for collisions at center-of-mass energies of 
$\sqrt{s} = 160$~GeV and 240~GeV using \MG~5 and \pythia~6 Monte Carlo simulations based on the EPA
approach. Realistic jet acceptance and reconstruction efficiencies are applied as well as selection criteria
to enhance the signals over the relevant backgrounds. In the case of the $W^{+}W^{-}$ analysis,
parton-showering and hadronization from \pythia~6  is combined with the Durham algorithm for clustering of the
hadronic energy into four exclusive jets.\\  

Observation of both processes at the 5$\sigma$-level is achievable for the expected
\FCCee\ luminosities. The measurement of $\gaga \to WW \to 4$~jets will yield more than 600
final counts which will allow for detailed studies of the trilinear $WW\gamma$ and quartic $\gaga WW$
couplings, either in the standard model or assuming new physics scenarios in terms of dimension-6 and 8 effective
operators~\cite{CMSaaww}. Extraction of limits on anomalous quartic gauge couplings from these simulations
requires more advanced phenomenological studies since higher-dimension effective operators using photon
fluxes depend strongly on new implementations in the MC generator tools.
A full simulation (including {\sc geant}-based detector response) of both processes would provide more
realistic conditions which are, however, beyond the scope of these proceedings and should not
change its main conclusions. 
The feasibility analyses developed in this work confirm the unique Higgs and electroweak physics potential open
to study in $\gaga$ collisions at the \FCCee. 

\ack
P. R. T. thanks Photon'15 and the CMS Collaboration for financial support. This work has granted partially by
the Brazilian Science without Borders Program from Coordination for the Improvement of Higher Education
Personnel -- CAPES (contract BEX 11767-13-8).


\end{document}